\documentstyle[a4]{article}
\begin{document}
\large
\title{Extended Born - Oppenheimer equation for tri - state system} 
\author{Biplab Sarkar and Satrajit Adhikari\footnote{
        Author to whom correspondence should be addressed: Fax: +91-361-690762;
        Electronic mail: satrajit@iitg.ernet.in}\\
	Department of Chemistry\\
	Indian Institute of Technology, Guwahati\\
	North Guwahati, Guwahati - 781 039\\
	INDIA}
\maketitle
	\begin{center}
	{\bf {ABSTRACT}}
	\end{center}

	\vspace*{0.5cm}
	
         \noindent
          We present explicit form of non - adiabatic coupling (NAC) elements of nuclear Schroedinger equation (SE) 
          for a coupled tri - state electronic manifold in terms of mixing angles of real electronic basis functions.
          If the adiabatic - diabatic transformation (ADT) angles are the mixing angles of electronic basis, ADT matrix 
          transforms NAC terms to exactly zeros and brings diabatic form of SE. ADT and NAC matrices satisfy a curl
          condition and find a novel relation among the mixing angles for irrotational case. We also find that 
          extended Born - Oppenheimer (EBO) equations have meaningful solution and can reproduce numerically exact 
          results only when the equations are gauge invariant. 

	\newpage

	\noindent
         Herzberg and Longuet - Higgins (HLH) [1] made an interesting observation in molecular dynamics - a real valued 
         electronic wavefunction changes sign when the nuclear coordinates encircles a closed path around a conical
         intersection. This is so - called geometric phase (GP) effect, where the total wavefunction is not well 
         behaved and the Born - Oppenheimer (BO) treatment [2] breaks down. HLH corrected this problem in an 
         {\it ad hoc} way by multiplying a complex phase factor with real double - valued electronic wavefunction
         such that the resulting complex electronic wavefunction is single valued. Mead and Truhlar [3] generalized the BO
         equation by introducing a vector potential into the nuclear Hamiltonian to account this effect and the 
         approach is reminiscent of HLH complex phase factor treatment. The effect has been found to show up
         immense significance, particularly, in scattering processes[4].

	\vspace*{0.3cm}
  
        \noindent
         The BO treatment is based on the fact that slow - moving nuclei can be distinguished from fast - moving electrons
         in molecular systems. This distinction leads to the BO approximation, which includes the assumption that NAC 
         terms are negligibly small and the upper electronic states do not affect the wavefunction on the lower state. 
         Though the implication of this assumption is considered to be independent on the energy of the system, the 
         ordinary BO equation was also applied for cases with large NAC terms assuming that upper electronic states are
         classically closed at low energies. Even if the components of total wavefunction on the upper electronic
         state(s) are negligibly small at enough low energies, their product with large NAC terms are definite values 
         and the BO approximation breaks down for those cases.                

	\vspace*{0.3cm}

        \noindent
         Since NAC terms appear as off - diagonal elements in the SE [5], formulations of EBO equation are the attempts [6-7]
         to include the effect of off - diagonal (non - adiabatic) elements on the single surface equation. Baer [7] derived
         a new set of potential energy coupled BO equations for two electronic states, where the effects of NAC terms 
         are being translated on the diagonal. At enough low energy, Baer assumed that upper electronic state is classically
         closed and approximate EBO equations for both the surfaces were formed. This EBO equation was used to calculate
         transition probabilities in a two - arrangement - channel model [8-9] and reproduced the correct results obtained 
         from rigorously derived diabatic SE. Varandas and Xu [10] expressed NAC elements of adiabatic nuclear SE in terms 
         of nuclear coordinate dependent mixing angle of two real electronic wavefunctions, found that ADT angle [11] is 
         identical (up to an additive constant) with the mixing angle and indeed, formulated single surface EBO equations 
         in the vicinity of degeneracy.

	\vspace*{0.3cm}

        \noindent
         Baer et al. [12-13] considered coupled BO equations for N ($\ge 2$) adiabatic nuclear SE and derived single surface 
         EBO equations for a model situation ( not a general case). These EBO equations contain the effect of 
         NAC terms where the ground adiabatic PES have a degeneracy with excited surfaces along a single line. Even 
         though this derivation was not persuaded for a general situation, the formulation shows that such an 
         extension is viable and those EBO equations will have meaningful solution only when they are gauge invariant.  
        
	\vspace*{0.3cm}

        \noindent
         In this article, we carry out the BO treatment of a coupled three state electronic manifold from first principles.
         The BO expansion of molecular wavefunction and the total electron - nuclei Hamiltonian in the adiabatic 
         representation are given by: 
 
        \begin{eqnarray}
        \Psi({\bf n},{\bf e}) &=& \sum_{i=1}^3\psi_i({\bf n}) \xi_i({\bf e},{\bf n}), \nonumber\\
        \hat{H} &=& \hat{T}_n + \hat{H}_e({\bf e},{\bf n}),\nonumber\\
        \hat{T}_n &=& -\frac{\hbar^2}{2m}\sum_n\nabla^2_n, \nonumber\\
        \hat{H}_e({\bf e},{\bf n})\xi_i({\bf e},{\bf n}) &=& u_i({\bf n}) \xi_i({\bf e},{\bf n}),
        \end{eqnarray}
                                                                                                                             
        \noindent
         where ${\bf n}$ and ${\bf e}$ are the sets of nuclear and electronic coordinates, respectively, 
         $\xi_i({\bf e},{\bf n})$ is the eigenfunction of the electronic Hamiltonian, $\hat{H}_e({\bf e},{\bf n})$ with 
         eigenvalue, $u_i({\bf n})$, also $\psi_i({\bf n})$ is the nuclear coordinate dependent expansion coefficient and 
         $\hat{T}_n$ is the nuclear kinetic energy (KE) operator. When we substitute equation (1) in the time - independent
         SE and multiply by electronic basis, $\xi_i({\bf e},{\bf n})$, the matrix representation of adiabatic nuclear 
         equation takes the following form after integrating over the electronic coordinates, ${\bf e}$, 
                                                                                                                             
        \begin{eqnarray}
         \sum_{j=1}^3(H_{ij}&-&E\delta_{ij})\psi_j({\bf n})=0, \hspace{0.3cm} i=1,2,3,\nonumber\\
           H_{ii} &=&-\frac{\hbar^2}{2m}(\nabla^2+2\vec{\tau}^{(1)}_{ii}{\bf \cdot}\vec{\nabla} 
           +\tau^{(2)}_{ii})+u_i({\bf n}),\nonumber\\
           H_{ij} &=&-\frac{\hbar^2}{2m}(2\vec{\tau}^{(1)}_{ij}{\bf \cdot}\vec{\nabla} +\tau^{(2)}_{ij})=H_{ji}^{\dagger},
           \nonumber\\
           \vec{\tau}^{(1)}_{ij} &=& \langle \xi_i ({\bf e},{\bf n})|\vec{\nabla} |\xi_j ({\bf e},{\bf n})\rangle, 
           \hspace{0.5cm}
           \tau^{(2)}_{ij} = \langle \xi_i ({\bf e},{\bf n})|\nabla^2 |\xi_j ({\bf e},{\bf n})\rangle,\nonumber\\ 
           & &\langle \xi_i({\bf e},{\bf n}) |\xi_j ({\bf e},{\bf n})\rangle = \delta_{ij}.
        \end{eqnarray}
                                                                                                                             
        \noindent
         We define the following three real orthonormal functions as the electronic basis set ($\xi_1$, $\xi_2$ and $\xi_3$)
         where $\alpha({\bf n})$, $\beta({\bf n})$ and $\gamma({\bf n})$ are the nuclear coordinate dependent mixing angles
         (e.g. $\alpha({\bf n})$ is the mixing angle between electronic states 1 and 2), 
                                                                                                                             
        \begin{eqnarray}
           \xi_1 &=& \left( \begin{array}{c}
           \cos\alpha \cos\beta \\
           \sin\alpha \cos\beta\\
           \sin\beta
           \end{array}\right), \hspace{0.5cm}
           \xi_2 = \left( \begin{array}{c}
           -\cos\alpha\sin\beta\sin\gamma-\sin\alpha\cos\gamma\\
           -\sin\alpha\sin\beta\sin\gamma+\cos\alpha\cos\gamma\\
            \cos\beta\sin\gamma
           \end{array}\right),\nonumber\\
           \xi_3 &=& \left( \begin{array}{c}
           -\cos\alpha\sin\beta\cos\gamma+\sin\alpha\sin\gamma\\
           -\sin\alpha\sin\beta\cos\gamma-\cos\alpha\sin\gamma\\
            \cos\beta\cos\gamma
           \end{array}\right),
        \end{eqnarray}
                                                                                                                             
        \noindent
        and rewrite the kinetically coupled nuclear SE (equation (2)) as below,

    \begin{eqnarray}
    -\frac{\hbar^2}{2m}\left( \begin{array}{ccc}
           \vec{\nabla} & \vec{t}_1 & \vec{t}_2\\
            -\vec{t}_1 & \vec{\nabla}& \vec{t}_3\\
            -\vec{t}_2 & -\vec{t}_3 & \vec{\nabla}\\
           \end{array}\right)^2\left( \begin{array}{l}
           \psi_1\\ \psi_2\\ \psi_3
           \end{array}\right)+
           \left( \begin{array}{ccc}
            u_1-E& 0 & 0\\
            0 & u_2-E & 0\\
            0 & 0 & u_3-E\\
           \end{array}\right)
           \left( \begin{array}{l}
           \psi_1\\ \psi_2\\ \psi_3
           \end{array}\right)= 0,
        \end{eqnarray}
                                                                                                                             
      \noindent
       where the NAC matrix ($\vec{\bf \tau} (\equiv \vec{\bf \tau}^{(1)})$) is defined as,

      \begin{eqnarray}
           \vec{\bf \tau}=\left( \begin{array}{ccc}
             0   & \vec{t}_1 & \vec{t}_2\\
            -\vec{t}_1 & 0   & \vec{t}_3\\
            -\vec{t}_2 & -\vec{t}_3 & 0\\
           \end{array}\right),
        \end{eqnarray}
                                                                                                                             
      \noindent
      with matrix elements, 
      
       \begin{eqnarray}
       \vec{t}_1&=&-\cos\beta\cos\gamma\vec{\nabla}\alpha - \sin\gamma\vec{\nabla}\beta,\nonumber\\
       \vec{t}_2&=&\cos\beta\sin\gamma\vec{\nabla}\alpha-\cos\gamma\vec{\nabla}\beta,\nonumber\\
       \vec{t}_3&=&-\sin\beta\vec{\nabla}\alpha-\vec{\nabla}\gamma.
       \end{eqnarray}

      \noindent
       When we substitute $\Psi={\bf A}\Psi^d$ in equation (4) with the following choice of ADT matrix,                                                                                                                                    
      \begin{eqnarray}
       {\bf A}=\left( \begin{array}{ccc}
                \cos\alpha\cos\beta & \sin\alpha\cos\beta & \sin\beta\\\\
               -\sin\alpha\cos\gamma & \cos\alpha\cos\gamma & \cos\beta\sin\gamma\\
               -\cos\alpha\sin\beta\sin\gamma& -\sin\alpha\sin\beta\sin\gamma& \\\\
               \sin\alpha\sin\gamma & -\cos\alpha\sin\gamma & \cos\beta\cos\gamma\\
               -\cos\alpha\sin\beta\cos\gamma& -\sin\alpha\sin\beta\cos\gamma& 
               \end{array}\right),
      \end{eqnarray}
                                                                                                                             
       \noindent
        adiabatic nuclear SE is being transformed to the potentially coupled diabatic nuclear SE,                                                                                                                              
        \begin{eqnarray}
           \sum_{j=1}^3\{(-\frac{\hbar^2}{2m}\nabla^2-E)\delta_{ij}&+&W_{ij}\}\psi_j^d= 0, \hspace{0.3cm} i=1,2,3,\nonumber\\
        {\bf W} &=& {\bf A}^{\dagger}{\bf U} {\bf A}, \hspace{0.5cm} U_{ij} = u_i\delta_{ij}.
        \end{eqnarray}

         \noindent
         Since we find that the above form of $\vec{\bf \tau}$ (equation (5) and (6)) and ${\bf A}$ (equation (7)) matrices satisfy
         the ADT condition [11],
 
         \begin{eqnarray}
         \vec{\nabla} {\bf A} + \vec{\bf \tau} {\bf A} = 0,
         \end{eqnarray}                                                                                                                              
                                                                                                                             
         \noindent
          we arrive an equation (known as curl condition) for each NAC element, $\vec{\tau}_{ij}$,
          considering the analyticity of the transformation matrix ${\bf A}$ for any two nuclear coordinates, $p$ and $q$,

         \noindent
         \begin{eqnarray}
         \frac{\partial}{\partial p} \tau_{ij}^q - \frac{\partial}{\partial q} \tau_{ij}^p 
          &=& {({\bf \tau}^q {\bf \tau}^p)}_{ij} - {({\bf \tau}^p {\bf \tau}^q)}_{ij},\nonumber\\
         \tau^p_{ij} = <\xi_i| \nabla_p|\xi_j>, &\hspace{0.3cm}& \tau^q_{ij} = <\xi_i| \nabla_q|\xi_j>.
          \end{eqnarray}

         \noindent
         The curl condition for each pair of electronic basis, $\{|\xi_1>, |\xi_2>\}$, 
         $\{|\xi_1>, |\xi_3>\}$ and $\{|\xi_2>, |\xi_3>\}$ is satisfied and the explicit forms of curl 
          equations are the following, 
       
         \begin{eqnarray}
         Curl\tau_{12}=[{\bf \tau}\times {\bf \tau}]_{12} &=&\sin\beta\cos\gamma[\nabla_p\alpha\nabla_q\beta 
         - \nabla_q\alpha\nabla_p\beta] 
         + \sin\gamma\cos\beta[\nabla_p\alpha\nabla_q\gamma-\nabla_q\alpha\nabla_p\gamma]\nonumber\\
         &-&\cos\gamma[\nabla_p\beta\nabla_q\gamma-\nabla_q\beta\nabla_p\gamma],\nonumber\\
         Curl\tau_{13}=[{\bf \tau}\times {\bf \tau}]_{13} &=&-\sin\beta\sin\gamma[\nabla_p\alpha\nabla_q\beta 
         - \nabla_q\alpha\nabla_p\beta]
         + \cos\gamma\cos\beta[\nabla_p\alpha\nabla_q\gamma-\nabla_q\alpha\nabla_p\gamma]\nonumber\\
         &+&\sin\gamma[\nabla_p\beta\nabla_q\gamma-\nabla_q\beta\nabla_p\gamma],\nonumber\\
         Curl\tau_{23}=[{\bf \tau}\times {\bf \tau}]_{23} &=&cos\beta[\nabla_p\alpha\nabla_q\gamma
         -\nabla_q\alpha\nabla_p\gamma].
         \end{eqnarray}
                                                                                                                             
         \noindent
          Since $\vec{\bf \tau}({\bf n})$ goes to zero rapidly enough as the radial coordinate tends to infinity [14], $\vec{\bf \tau}$
          may be resolved into an irrotational and a solenoidal part [15-16]. On the otherhand, the explicit form of
          div$\tau_{ij}$s are given by:

         \begin{eqnarray}
          div \vec{\tau}_{12}&=&2\sin\beta\cos\beta\sin\gamma(\nabla\alpha)^2+3\cos\beta\sin\gamma\nabla\alpha\nabla\gamma
          -3\cos\gamma\nabla\beta\nabla\gamma-\cos\beta\cos\gamma\nabla^2\alpha\nonumber\\ &-&\sin\gamma\nabla^2\beta
          -\sin\beta\cos\gamma\nabla\alpha\nabla\beta\nonumber\\
          div \vec{\tau}_{13}&=&2\sin\beta\cos\beta\cos\gamma(\nabla\alpha)^2+3\cos\beta\cos\gamma\nabla\alpha\nabla\gamma
          +3\sin\gamma\nabla\beta\nabla\gamma+\cos\beta\sin\gamma\nabla^2\alpha\nonumber\\ &-&\cos\gamma\nabla^2\beta
          +\sin\beta\sin\gamma\nabla\alpha\nabla\beta\nonumber\\
          div\vec{\tau}_{23}&=&2\cos^2\beta\sin\gamma\cos\gamma(\nabla\alpha)^2-2\sin\gamma\cos\gamma(\nabla\beta)^2
          -3\cos\beta\cos^2\gamma\nabla\alpha\nabla\beta \nonumber\\ &+& \cos\beta\sin^2\gamma\nabla\alpha\nabla\beta
          - \sin\beta\nabla^2\alpha - \nabla^2\gamma
         \end{eqnarray}

         \noindent
          Since $\nabla\alpha$, $\nabla\beta$ and $\nabla\gamma$ (generally $\nabla^2\alpha$, $\nabla^2\beta$ and 
          $\nabla^2\gamma$ also) are not zero in the vicinity of the conical intersection, div$\vec{\tau}_{ij}\ne 0$ 
          for any value of mixing angles, i.e., vector field $\vec{\bf \tau}$
          corresponds to non - solenoidal case[16-17]. In the vicinity of the conical intersection, we 
          presently handle only the irrotational part, i.e., curl equations are Abelian, $Curl \tau_{ij} = 0$. 
          Thus, equations (11) have unique solution as below (if $\beta$ $\ne$ $\frac{\pi}{2}$ 
          or $\ne$ $\frac{3\pi}{2}$),
                                                                                                                             
         \begin{eqnarray}
         \nabla_p \alpha \nabla_q \beta &=& \nabla_q\alpha\nabla_p\beta,\nonumber\\
         \nabla_p \beta \nabla_q \gamma &=& \nabla_q\beta\nabla_p\gamma,\nonumber\\
         \nabla_p \alpha \nabla_q \gamma &=& \nabla_q\alpha\nabla_p\gamma,
         \end{eqnarray}
                                                                                                                             
         \noindent
          with the implication that nuclear coordinate dependent mixing angles are related by integer ratios, 
                                                                                                                             
        \begin{eqnarray}
        \alpha({\bf n}):\beta({\bf n}):\gamma({\bf n}) = k:l:m.
        \end{eqnarray}
                                                                                                                             
        \noindent
        If an unitary transformation matrix, ${\bf G}$ $(\Psi = {\bf G}\Phi)$, diagonalizes NAC 
        matrix, $\vec{\bf \tau}$, with eigenvalues, 0 and $\pm i\vec{\omega}$, the adiabatic SE (equation (4)) transforms as, 
                                                                                                                             
        \begin{eqnarray}
        -\frac{\hbar^2}{2m}(\vec{\nabla}&+&i\vec{\omega})^2\Phi + ({\bf V}-E)\Phi = 0, 
         \hspace{0.3cm} {\bf V}={\bf G}^{\star} {\bf U} {\bf G}, \nonumber\\
        \vec{\omega}&=&\pm\sqrt{t_1^2+t_2^2+t_3^2}\nonumber\\
        &=&\pm\{(\vec{\nabla} \alpha)^2 + (\vec{\nabla} \beta)^2 + (\vec{\nabla} \gamma)^2
        + 2 \sin\beta \vec{\nabla}\alpha \vec{\nabla}\gamma\}^{\frac{1}{2}}.
        \end{eqnarray}
                                                                                                                             
        \noindent
         One can rewrite the product, ${\bf V}\Phi$, for the $i${\it th} equation as,
         $(V\Phi)_i = u_1\Phi_i + \sum_{j=2}^3 G^{\star}_{ij}(u_i-u_1)\psi_j$, $i=1,2,3$ and impose the
         BO approximation,  $|\psi_1| >> |\psi_i|$, $i=2,3$ (considering that at enough low energy, both the
         upper electronic states are assumed to be classically closed) to form the
         single surface adiabatic nuclear SE [12],

         \begin{eqnarray}
         -\frac{\hbar^2}{2m}(\nabla+i\omega_i)^2\Phi_i + (u_1-E)\Phi_i=0, \hspace{0.5cm} i=1,2,3.
         \end{eqnarray}
                                                                                                                             
         \noindent
         Equation (14) simplifies both the adiabatic and diabatic equations (4) and (8), respectively for any arbitrary 
         ratios of mixing angles, i.e., the NAC matrix takes the form, 
         $\vec{\bf \tau}=\vec{\nabla}\alpha.{\bf g} (\alpha)$, where ${\bf g}(\alpha)$ is mixing angle dependent 
         $3\times 3$ matrix. At the same time,  we know that EBO equations (16) have meaningful solution only when 
         they satisfy the following gauge invariance condition [12] for systems of three electronic states,
                                                                                                                             
        \begin{eqnarray}
        \frac{1}{2\pi} \int_0^{2\pi} \vec{\omega(\bf n)}.\vec{d\bf n} = m, \hspace{0.3cm} m=1,2,3,.... 
        \end{eqnarray}

        \noindent
         We choose different ratios of $\alpha(\bf n)$, 
         $\beta(\bf n)$ and $\gamma(\bf n)$ and calculate corresponding $\omega$ s as: (a) $\alpha$ = $\beta$ = $\gamma$,
         \hspace{0.5cm} $\vec{\omega} =\pm\vec{\nabla}\alpha\{3+ 2 \sin\alpha\}^{\frac{1}{2}}$; \hspace{0.5cm} 
         (b) $\alpha$ = $2\beta$ = $\gamma$, \hspace{0.5cm} $\vec{\omega}=\pm\frac{\vec{\nabla}\alpha}{2}
         \{9+ 8 \sin\frac{\alpha}{2}\}^ {\frac{1}{2}}$; \hspace{0.5cm} (c) $2\alpha$ = $\beta$ = $2\gamma$, 
         \hspace{0.5cm} $\vec{\omega}= \pm\vec{\nabla}\alpha \{6+ 2 \sin2\alpha\}^{\frac{1}{2}}$ where $\alpha(\bf n)$ 
         is mixing as well as ADT angle (upto an additive constant) among the electronic states. 
         It is important to note that in all such situations, divergence equations (12) have non - zero contributions.  
         When $\alpha(\bf n)$
         is the function of two nuclear coordinates, $x(=q\cos\theta)$ and $y(=q\sin\theta)$and is being equated as 
         $\alpha(\bf n)=\frac{\theta}{2}$, the product, $\vec{\nabla}\alpha({\bf n}) \cdot \vec{d {\bf n}} = 
         \frac{1}{2} d\theta$. For realistic systems, $\alpha(\theta)$ can be calculated as function of $\theta$ from 
         the electronic eigenfunctions of the equation, $ \hat{H}_e({\bf e},{\bf n})\xi_i({\bf e},{\bf n}) = 
         u_i({\bf n}) \xi_i({\bf e},{\bf n})$. Thus, the gauge invariant integrals for the above three cases are, 
                                                                                                                             
         \begin{eqnarray}
         \Gamma_1=\frac{1}{2\pi} \int_0^{2\pi} \frac{1}{2} \{3+ 2 \sin\frac{\theta}{2}\}^{\frac{1}{2}} d\theta
                           &=& 4\sqrt{5}\int_{0}^{\frac{\pi}{4}} (1-\frac{4}{5}\sin^2\phi)^{\frac{1}{2}} d\phi
                                                            = 1.03\\
         \Gamma_2=\frac{1}{2\pi} \int_0^{2\pi} \frac{1}{4} \{9+ 8 \sin\frac{\theta}{4}\}^{\frac{1}{2}} d\theta
                           &=& 2\sqrt{17}\int_{0}^{\frac{\pi}{4}} (1-\frac{16}{17}\sin^2\phi)^{\frac{1}{2}} d\phi
                                                            = 0.934\\
         \Gamma_3=\frac{1}{2\pi} \int_0^{2\pi} \frac{1}{2} \{6+ 2 \sin\theta\}^{\frac{1}{2}} d\theta
                           &=& 2\sqrt{2}\int_{0}^{\frac{\pi}{4}} (1-\frac{1}{2}\sin^2\phi)^{\frac{1}{2}} d\phi\nonumber\\
                           &+& 2\sqrt{2}\int_{0}^{\frac{3\pi}{4}} (1-\frac{1}{2}\sin^2\phi)^{\frac{1}{2}} d\phi=1.216
         \end{eqnarray}
                                                                                                                             
         \noindent
          with the general form of {\it incomplete elliptic integral of the second kind}. It is quite obvious that since 
          $\alpha(\bf n)$ ({\bf A} is analytic) is analytic, the nature of these integrals will be generic for 
          any functional form of $\alpha({\bf n})$. When three electronic states are coupled, the non - adiabatic effect
          of the upper states on the ground is equivalent to a potential developed due to elliptic motion of the nuclei
          around the point of conical intersection. Moreover, single surface EBO can be derived 
          only for specific ratios of mixing/ADT angles, e.g., gauge invariance condition is approximately
          obeyed in case (a), whereas in the cases (b) and (c), integrals are either away or far away from the gauge
          condition.

	\vspace*{0.3cm}

         \noindent
          Since the general form of $\vec{\bf \tau}$ and ${\bf A}$ (equations (5) - (7)) with any arbitrary ratios of mixing angles
          satisfies the equation, $\vec{\nabla} {\bf A} + \vec{\bf \tau} {\bf A} = 0$, and ensures the ADT, uniquely defined diabatic
          potential matrix in configuration space is guaranteed by the unit matrix, ${\bf D}=\exp(\int_0^{2\pi} \vec{\bf \tau} \cdot 
          \vec{d\bf n})$. The explicit expression of ${\bf D}$ [13] is derived for the case (a) by using the corresponding 
          ${\bf G}$ matrix (equation (15)) as,
          
          \begin{eqnarray}
          {\bf D} &=& {\bf G} \exp(-i\int_0^{2\pi}\vec{\omega}({\bf n})\cdot \vec{d{\bf n}}) {\bf G}^{\star}=\frac{1}{3+2.si}\nonumber\\                \nonumber\\
            &\times& 
           \left( \begin{array}{ccc}
            (si+1)^2                    &  -(cs^2+si)                        &(cs.si-cs)\\
            +(1+cs^2)C_1                & \times (3+2.si)^{\frac{1}{2}}S_1   &\times (3+2.si)^{\frac{1}{2}}S_1\\
                                        &   +2.cs^3S_2                       &+2(1+si+si.cs^2)S_2\\\\          
            -(cs^2+si)                  &(cs.si-cs)^2                        & -(si+1)(3+2.si)^{\frac{1}{2}}S_1\\
            \times (3+2si)^{\frac{1}{2}}S_1 &(1+2si+2.si^2                        & +2.cs(si^3+cs^2-si^2)S_2\\ 
            +2cs^3 S_2                  &+2.si.cs^2+cs^4)C_1                  &\\\\
            -(cs.si-cs)                 & (si+1)(3+2.si)^{\frac{1}{2}}S_1    & (cs^2+si)^2\\
            \times (3+2.si)^{\frac{1}{2}}S_1 & +2.cs(si^3+cs^2-si^2)S_2       & + (2+2.si^3\\
             +2(1+si+si.cs^2))S_2        &                                    & +si^2+cs^2)C_1\\
           \end{array}\right)
            \nonumber\\
            &\simeq& \left( \begin{array}{ccc}
                1   & 0 & 0\\ 
                0   & 1 & 0\\ 
                0   & 0 & 1\\ 
           \end{array}\right)
          \end{eqnarray}
            
           \noindent
            where $si=\sin\alpha$, $cs=\cos\alpha$, $C_1=\cos(2\pi\Gamma_1)\simeq 1$, $S_1=\sin(2\pi\Gamma_1)\simeq 0$ and
            $S_2=\sin^2(\pi\Gamma_1)\simeq 0$ and for the other two cases (b) and (c), $C_1\neq 1$ and $S_1, S_2 \neq 0$. 

	\vspace*{0.3cm}

           \noindent
            In equation (15), the contribution of non - adiabatic effects appear as $\pm i\omega$ in the KE 
            operator (diagonal) as well as in the potential energy matrix (through ${\bf G}$ matrix). Since single surface 
            EBO equation is derived by neglecting the effect of ${\bf G}$ matrix ($\sum_{j=2}^3 G^{\star}_{ij} (u_i-u_1)\psi_j$), 
            diabatic and single surface EBO equations will not be equivalent irrespective of any arbitrary ratios of mixing
            angles. Alternatively, we can say that for any arbitrary ratios of mixing angles, the term,
            $\sum_{j=2}^3 G^{\star}_{ij} (u_i-u_1)\psi_j$, is not negligible and single surface EBO equations can not
            be derived.  The condition of gauge invariance and the existence of uniquely define diabatic potential from 
            non - adiabatic terms clearly tell that only for specific ratios of mixing angles, single surface EBO 
           equations is acceptable. 
     
	\vspace*{0.3cm}

          \noindent
           We justify our theory by performing numerical calculations on a two coordinate quasi -``JT scattering'' model
           [13] in which the harmonic oscillator potential and linear coupling term are replaced by more general potentials.           The adiabatic PESs $u_l$, l = 1,2,3 are given below:
                                                                                                                             
         \begin{eqnarray}
          u_1(x,y) &=& \frac{1}{2}\mu (\omega_0-\tilde{\omega}_1(x))^2 y^2 + A_1 \times f(x,y)\nonumber\\
          u_2(x,y) &=& \frac{1}{2}\mu \omega_0^2 y^2 - (D_1-A_1) \times f(x,y)+D_1\nonumber\\
          u_3(x,y) &=& \frac{1}{2}\mu \omega_0^2 y^2 - (D_2-A_1) \times f(x,y)+D_2\nonumber\\
         \tilde{\omega}_1(x) &=& \omega_1 \exp((-\frac{x}{\sigma_1})^2)\nonumber\\
          f(x,y) &=& \exp\Big( - \frac{x^2 + y^2}{\sigma^2}\Big)
          \end{eqnarray}
                                                                                                                             
         \noindent
          where $\mu=0.58$ $amu$, $A_1=3.0$ $eV$, $D_1=5.0$ $eV$, $D_2=10.0$ $eV$, $\omega_0=39.14\times 10^{13}$ $s^{-1}$,
          $\omega_1=7.83\times 10^{13}$ $s^{-1}$, $\sigma=0.3$ $\AA$ and $\sigma_1=0.75$ $\AA$, also Cartesian coordinates
          $x$ and $y$ are defined in the intervals $-\infty \le x \le \infty$ and $-\infty \le y \le \infty$ and 
          related with polar coordinates as, $x=q\cos\theta$ and $y=q\sin\theta$. These adiabatic potentials
          describe a two arrangement channel system where $x \rightarrow \infty$ and $x \rightarrow -\infty$ are
          the reagents and products asymptote, respectively. 

	\vspace*{0.3cm}
         
        \noindent
         Introducing $\alpha({\bf n}) = \frac{\theta}{2} = \frac{1}{2}\tan^{-1}(\frac{y}{x})$, we 
         construct diabatic potential matrices (${\bf W}$), vis -a- vis, diabatic and single surface EBO equations for all
         the three cases and then, initialize the wavefunction on the ground vibrational state with different initial KE
         at the asymptote of reagents, propagate the time dependent wavefunction using DVR [18] and project the final 
         wavefunction with the asymptotic eigenfunctions of the Hamiltonian to calculate state - to - state vibrational
         transition probabilities. Dynamical calculations are carried out at total energy 1.20 and 1.80 eV. Since the 
         point of conical intersection is at 3.0 eV, upper electronic states are expected to be classically closed at
         those energies. We demonstrate all the results in Table (I) - (III) for the cases (a) - (c), respectively. 
         Table(I) exhibits that reactive transition probabilities calculated by single surface EBO equations not only 
         follow the correct symmetry but also achieve quantitative agreement with diabatic results whereas in Table (II)
         and (III), single surface EBO results are symmetry broken as well as inaccurate. Thus, these numerical results
         predict that only for the case (a), single surface EBO equation can be constructed.
          
	\vspace*{0.3cm}

        \noindent
         In summary, we have used the generalized form of real orthogonal electronic basis functions in terms of mixing 
         angles ($\alpha({\bf n})$, $\beta({\bf n})$ and $\gamma({\bf n})$) among the three electronic states and expressed 
         the NAC terms of adiabatic nuclear SEs with these angles. When ADT angles are chosen as mixing
         angles, adiabatic nuclear SEs transform to diabatic SEs, i.e, ADT and mixing angles are equal upto an 
         additive constant. Since the NAC ($\vec{\bf \tau}$), ADT (${\bf A}$) matrices satisfy the ADT condition
         and mixing angles are analytic, we find a curl condition is also satisfied with non - zero divergence 
         for each element of $\vec{\bf \tau}$ 
         matrix. In irrotational case, solution of these curl equations imply that mixing/ADT angles are related 
         with integer ratios. Consequently, both the adiabatic and diabatic nuclear SEs are being simplified extensively. 
         Single surface EBO equations derived from adiabatic nuclear SEs have meaningful solution
         and quantitative agreement with corresponding diabatic case only for specific ratios of mixing angles where EBO 
         equation is gauge invariant and provide uniquely defined diabatic potential energy matrix. We also find that 
         the non - adiabatic effect associated with single surface EBO equation derived from three coupled electronic 
         state is a potential that arises due to the elliptic motion of nuclei around the conical intersection.  
          
	\vspace*{0.3cm}

        \noindent
         We acknowledge {\bf Department of Science and Technology (DST, Government of India)} for financial support 
         through the project no. {\bf SP/S1/H-53/01}. S.A. would like to thank Professor S. P. Bhattacharyya and
         Professor J. K. Bhattacharjee, I.A.C.S., Kolkata for their comments on curl and divergence equations.

         \newpage

         \noindent
         TABLE I

         \noindent
          Reactive state - to - state transition probabilities. Three diabatic surfaces are constructed considering 
          the relation, $\alpha({\bf n})=\beta({\bf n})=\gamma({\bf n})$ where the EBO is derived under the same 
          situation.
                                                                                                                             
         \vspace*{1.0cm}
                                                                                                                             
          \begin{tabular}{lllllllll} \hline \hline
           E (eV) & 0 $\rightarrow$ 0 & 0 $\rightarrow$ 1& 0 $\rightarrow$ 2&
           0 $\rightarrow$ 3 & 0 $\rightarrow$ 4& 0 $\rightarrow$ 5& 0 $\rightarrow$ 6\\
          \hline
           1.20 & 0.0229$^a$ & 0.0000 & 0.0616 & 0.0000 & 0.0006 & & \\
           1.20 & 0.0279$^b$ & 0.0032 & 0.0656 & 0.0049 & 0.0010 & & \\\\
           1.80 & 0.1002 & 0.0000 & 0.0401 & 0.0000 & 0.0912 & 0.0000 & 0.0238\\
           1.80 & 0.1194 & 0.0069 & 0.0524 & 0.0016 & 0.1196 & 0.0093 & 0.0139\\
           \hline\hline
           \end{tabular}
                                                                                                                             
           \vspace*{0.5cm}
                                                                                                                             
           \noindent
           $^a$ Diabatic\\
           $^b$ EBO

           \newpage
                                                                                                                             
            \noindent
             TABLE II

             \noindent
            Reactive state - to - state transition probabilities. Three diabatic surfaces are constructed considering
            the relation, $\alpha({\bf n})=2\beta({\bf n})=\gamma({\bf n})$ where the EBO is derived under the 
            same situation.

            \vspace*{1.0cm}
                                                                                                                             
            \begin{tabular}{lllllllll} \hline \hline
             E (eV) & 0 $\rightarrow$ 0 & 0 $\rightarrow$ 1& 0 $\rightarrow$ 2&
             0 $\rightarrow$ 3 & 0 $\rightarrow$ 4& 0 $\rightarrow$ 5& 0 $\rightarrow$ 6\\
             \hline
             1.20 & 0.0225$^a$ & 0.0000 & 0.0491 & 0.0000 & 0.0124 & & \\
             1.20 & 0.0114$^b$ & 0.0321 & 0.0298 & 0.0230 & 0.0002 & & \\\\
             1.80 & 0.1389 & 0.0001 & 0.0246 & 0.0000 & 0.0876 & 0.0000 & 0.0103\\
             1.80 & 0.0732 & 0.0684 & 0.0389 & 0.0036 & 0.0835 & 0.0190 & 0.0129\\
             \hline\hline
             \end{tabular}
                                                                                                                             
             \vspace*{0.5cm}
                                                                                                                             
             \noindent
             $^a$ Diabatic\\
             $^b$ EBO

           \newpage
                                                                                                                             
            \noindent
             TABLE III

             \noindent
            Reactive state - to - state transition probabilities. Three diabatic surfaces are constructed considering
            the relation, $2\alpha({\bf n})=\beta({\bf n})=2\gamma({\bf n})$ where the EBO is derived under the 
            same situation.

            \vspace*{1.0cm}
                                                                                                                             
            \begin{tabular}{lllllllll} \hline \hline
             E (eV) & 0 $\rightarrow$ 0 & 0 $\rightarrow$ 1& 0 $\rightarrow$ 2&
             0 $\rightarrow$ 3 & 0 $\rightarrow$ 4& 0 $\rightarrow$ 5& 0 $\rightarrow$ 6\\
             \hline
             1.20 & 0.0665$^a$ & 0.0000 & 0.0115 & 0.0000 & 0.0008 & & \\
             1.20 & 0.0315$^b$ & 0.0102 & 0.0629 & 0.0002 & 0.001 & & \\\\
             1.80 & 0.1220 & 0.0000 & 0.0390 & 0.0000 & 0.0697 & 0.0000 & 0.0027\\
             1.80 & 0.1318 & 0.0309 & 0.0363 & 0.0064 & 0.0586 & 0.0573 & 0.0061\\
             \hline\hline
             \end{tabular}
                                                                                                                             
             \vspace*{0.5cm}
                                                                                                                             
             \noindent
             $^a$ Diabatic\\
             $^b$ EBO

\newpage 

\noindent
{\bf References}

\begin{enumerate}

\item G. Herzberg and H. C. Longuet - Higgins, Discuss. Faraday Soc. {\bf 35}, 77 (1963). 
\item M. Born and J. R. Oppenheimer, Ann. Phys. (Leipzig) {\bf 84}, 457 (1927).
\item C. A. Mead and D. G. Truhlar, J. Chem. Phys. {\bf 70}, 2284 (1979).
\item A. Kuppermann and Y. -S. M. Wu, Chem. Phys. Lett. {\bf 205}, 577 (1993).
\item M. Baer, in {\it Theory of Chemical Reaction Dynamics}, edited by M. Baer
      (CRC Press, Boca Raton, FL, 1985), Vol. II, Chap. 4.
\item M. Baer and R. Englman, Chem. Phys. Lett. {\bf 265}, 105 (1996).
\item M. Baer, J. Chem. Phys. {\bf 107}, 10662 (1997). 
\item R. Baer, D. Charutz, R. Kosloff, and M. Baer, J. Chem. Phys. {\bf 105}, 9141 (1996).
\item S. Adhikari and G. D. Billing, J. Chem. Phys. {\bf 111}, 40 (1999).
\item A. J. C. Varandas and Z. R. Xu, J. Chem. Phys. {\bf 112}, 2121 (2000).
\item M. Baer, Chem. Phys. Lett. {\bf 35}, 112 (1975).
\item M. Baer, S. H. Lin, A. Alijah, S. Adhikari, and G. D. Billing, Phys. Rev. A {\bf 62}, 32506:1-8 (2000).
\item S. Adhikari, G. D. Billing, A. Alijah, S. H. Lin, and M. Baer, Phys. Rev. A {\bf 62}, 32507:1-7 (2000).
\item D. R. Yarkony, J. Chem. Phys. {\bf 84}, 3206 (1986).
\item G. B. Arfken, H. J. Weber, {\it Mathematical Methods for Physicists} (Academic Press Inc., San Diego, USA, 1995),
      Chap. 1.
\item D. J. Griffiths, {\it Introduction to Electrodynamics} (Printice-Hall, Inc., Englewood Cliffs, N.J., USA, 1989),
      Chap. 1. 
\item If infinitely long contour lines (seams) due to conical intersection are considered as infinitesimal narrow solenoids,
      experiments predict that seams should produce zero field outside of them but $\vec{\tau}$ differs
      from zero in the space surrounding the seams. 
\item P. Puzari, B. Sarkar, and S. Adhikari, J. Chem. Phys. {\bf 121}, 707 (2004). 
\end{enumerate}
\end{document}